\documentclass[aps,prb,twocolumn,showpacs,floatfix]{revtex4}
\usepackage{amssymb}
\usepackage{amsmath}
\usepackage{graphicx}

\begin{document}

\title{Dirac points of electron energy spectrum, band-contact
lines, and electron topological transitions of $3\frac{1}{2}$ kind
in three-dimensional metals}

\author{G.~P.~Mikitik}
\affiliation{B.~Verkin Institute for Low Temperature Physics \&
Engineering, Ukrainian Academy of Sciences, Kharkov 61103,
Ukraine}

\author{Yu.~V.~Sharlai}
\affiliation{B.~Verkin Institute for Low Temperature Physics \&
Engineering, Ukrainian Academy of Sciences, Kharkov 61103,
Ukraine}

\begin{abstract}
In many of three-dimensional metals with the inversion symmetry
and a weak spin-orbit interaction, Dirac points of the electron
energy spectrum form band-contact lines in the Brillouin zones of
these crystals, and electron topological transitions of
$3\frac{1}{2}$ kind are due to certain points of these lines. We
theoretically study these transitions in detail, and point out
that they can be detected with the magnetic susceptibility which
exhibits a giant diamagnetic anomaly at the $3\frac{1}{2}$-order
transitions.
\end{abstract}


\maketitle

\section{Introduction}

The Dirac points of electron energy spectra specify many unusual
properties of graphene, \cite{neto,sarma} topological insulators,
\cite{hasan,Qi,moore} and of Weyl semimetals. \cite{wan,balents}
In this paper we call attention to the fact that in ``usual''
three-dimensional (3D) metals the Dirac points can be located
along band-contact lines in the Brillouin zones of the metals, the
energies of the points can fill a whole interval in the energy
axis, \cite{c1} and these Dirac points lead to specific electron
topological transitions in the 3D metals. At these transitions
topology of the Fermi surface (FS) in the metals changes, but this
change differs from the FS transformation at the well-known
topological transitions of $2\frac{1}{2}$ kind, which were first
considered by Lifshits. \cite{lif} At the topological transitions
of $2\frac{1}{2}$ kind a new void of the FS appears (disappears),
or a neck of the FS forms (disrupts). These transitions occur near
critical points ${\bf p}_c$ in the Brillouin zone at which the
electron energy $\varepsilon$ has an extremum or a saddle point in
its dependence on the quasi-momentum ${\bf p}$. The small
vicinities of these points give singular van Hove contribution
\cite{vanhove} $\delta \nu(\varepsilon_F)$ to the electron density
of states $\nu(\varepsilon_F)$, $\delta \nu \propto (\varepsilon_F
-\varepsilon_c)^{1/2}$, where $\varepsilon_F$ is the Fermi energy
of the metal, and $\varepsilon_c=\varepsilon({\bf p}_c)$ is the
critical energy. In the $\Omega$-potential the appropriate
singular contribution has the form, $\delta
\Omega(\varepsilon_F)\propto (\varepsilon_F
-\varepsilon_c)^{5/2}$, and by analogy with the second-order
transitions, the topological transitions were named the
$2\frac{1}{2}$-order transitions although they are not due to a
change of any symmetry. The Lifshits topological transitions and
their manifestations in various physical phenomena were studied
theoretically and experimentally in many papers; see the reviews,
\cite{var,var1} the  recent papers, \cite{oka,hod} and references
therein. In this paper we show that the topological transitions
associated with the Dirac points are widespread in metals no less
than the $2\frac{1}{2}$-order transitions of Lifshits. The
transitions discussed here occur at the points of the
band-contact lines at which the energy of two touching bands
reaches its maximum or minimum on these lines. These transitions
are $3\frac{1}{2}$ kind according to the classification of Ref.
\onlinecite{lif}, and we explain how these transitions can be
detected in experiments.

\section{Band-contact lines}

It is common knowledge that the contact of the electron energy
bands in a metal can occur at symmetry points and along symmetry
axis of its Brillouin zone. Besides, as was shown by Herring,
\cite{herring} there are {\it lines of an accidental contact}
between two bands in crystals. The term "accidental" means that
the degeneracy of electron states is not caused by their symmetry.
Such band-contact lines along which the spectrum of the two bands
has the Dirac form are widespread in metals with the inversion
symmetry and with a weak spin-orbit interaction. This follows from
one of Herring's results: \cite{herring} if there is a point of an
intersection of two energy bands in an axis of symmetry of the
Brillouin zone, a band-contact line perpendicular to the axis has
to pass through this point. Intersection of the bands at points in
the axes frequently occurs even in simple metals. \cite{band-str}
The literature data show that the lines of the accidental contact
exist, for example, in Be, Mg, Zn, Cd, Al, graphite and many other
metals. The result of Herring can be understood from the following
simple considerations: \cite{lak} Let the two electron bands
marked by the indexes $1$ and $2$ be degenerate at a
quasi-momentum ${\bf p}_0$ and hence have the same energy
$\varepsilon_0$ at this point. The ${\bf k}\cdot{\bf p}$
Hamiltonian $\hat H$ for these bands in the vicinity of ${\bf
p}_0$ has the form:
 \begin{eqnarray}\label{1}
\hat H=\left (\begin{array}{cc} E_{11} & E_{12} \\ E_{12} & E_{22}
\end{array} \right),
 \end{eqnarray}
where $E_{12}={\bf v}_{12}\cdot({\bf p-p}_0)$,
$E_{11}=\varepsilon_0+{\bf v}_{11}\cdot({\bf p-p}_0)$,
$E_{22}=\varepsilon_0+{\bf v}_{22}\cdot({\bf p-p}_0)$, and ${\bf
v}_{ij}$ are the matrix elements of the velocity operator for the
electron states $1$ and $2$ at the point ${\bf p}_0$. Note that
${\bf v}_{12}$ and $E_{12}$ are real quantities for metals with
inversion symmetry and with negligible spin-orbit interaction. The
diagonalization of the Hamiltonian (\ref{1}) gives the dispersion
relation for the two bands in the vicinity of the point ${\bf
p}_0$:
 \begin{eqnarray}\label{2}
\varepsilon_{1,2}({\bf p})&=&\varepsilon_0 +{\bf v}_+ \cdot({\bf
p-p}_0) \\ \nonumber
&\pm& \left[\left({\bf v}_-\cdot({\bf
p-p}_0)\right)^2+ \left({\bf v}_{12}\cdot({\bf p-p}_0)\right)^2
\right]^{1/2},
 \end{eqnarray}
where ${\bf v}_+\equiv ({\bf v}_{11}+{\bf v}_{22})/2$, ${\bf v}_-
\equiv ({\bf v}_{11}-{\bf v}_{22})/2$. \cite{c2} It follows from
Eq.~(\ref{2}) that the spectrum has the Dirac form in the plane
passing through the vectors ${\bf v}_-$ and ${\bf v}_{12}$, and
the degeneracy of the bands $1$ and $2$ is not lifted along the
straight line passing through the point ${\bf p}_0$ in the
direction of the vector ${\bf z}=[{\bf v}_{-}\times {\bf v}_{12}]$
which is perpendicular to the plane. In other words, existence of
the band-degeneracy point leads to existence of the band-contact
line passing  through this point. Of course, this degeneracy line
exists not only near ${\bf p}_0$. If one takes into account terms
of higher order in ${\bf p-p}_0$ in the expansions of $E_{ij}$,
the band-contact line will be determined by the two equations:
$E_{11}({\bf p})-E_{22}({\bf p})=0$, $E_{12}({\bf p})=0$. Such
band-contact lines are either closed curves in the Brillouin zone
or end on the surface of this zone. In both the cases the energy
$\varepsilon_0$ of the two touching bands in the degeneracy line
is a periodic function of ${\bf p}_0$ running this line, and
$\varepsilon_0$ changes between its minimum $\varepsilon_{min}$
and maximum $\varepsilon_{max}$ values. In experiments, the
band-contact lines in metals can be, in principle, detected via
the phase analysis of the de Haas-van Alphen (or Shubnikikov-de
Haas) oscillations since the phase of these oscillations is
determined by the number of the band-contact lines penetrating the
appropriate cross-section of the Fermi surface.
\cite{jetp,prl,shen}

\begin{figure}[t] % %%%%%%%%%%%%%%%%%%%%%%%%%%%%%%%%%%%%%
 \centering  \vspace{+9 pt}
\includegraphics[scale=.80]{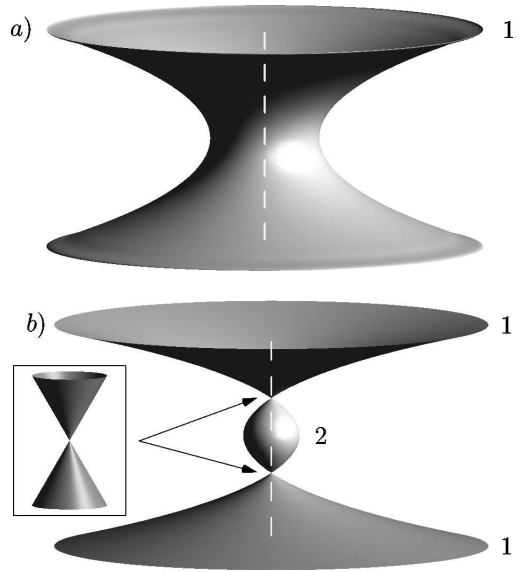}
\caption{\label{fig1} The Fermi surfaces near
$\varepsilon_c=\varepsilon_{min}$ at $\tilde a^2<1$ (here $\tilde
a=0$): (a) $\varepsilon_F < \varepsilon_{min}$, (b) $\varepsilon_F
> \varepsilon_{min}$. The band-contact line is shown by the dash
line. If $\varepsilon_c-\varepsilon_{min}$ increases, the two
conical points in the panel b) become widely separated, and the
inset shows the FS near one of these points when
$\varepsilon_{min}< \varepsilon_F <\varepsilon_{max}$, and
$\varepsilon_F$ is far away from these critical energies.
 } \end{figure}

\begin{figure}[t] % %%%%%%%%%%%%%%%%%%%%%%%%%%%%%%%%%%%%%
 \centering  \vspace{+9 pt}
\includegraphics[scale=.80]{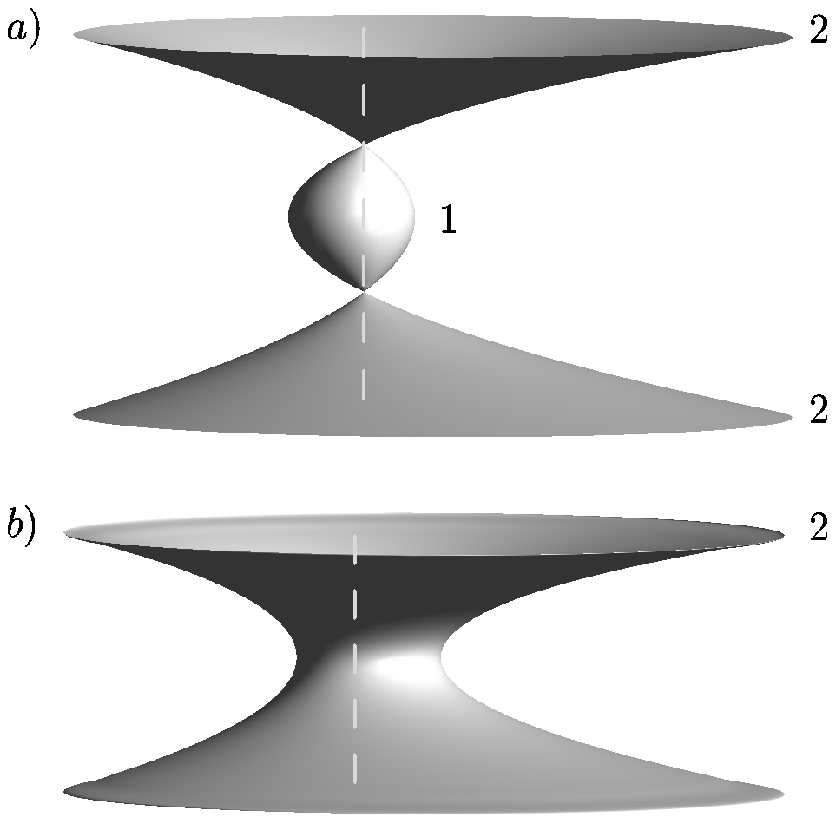}
\caption{\label{fig2} The Fermi surfaces near
$\varepsilon_c=\varepsilon_{max}$ at $0<\tilde a^2<1$: (a)
$\varepsilon_F < \varepsilon_{max}$, (b) $\varepsilon_F >
\varepsilon_{max}$. The band-contact line is shown by the dash
line.
 } \end{figure}

Let the Fermi level $\varepsilon_F$ of the metal lie in the
interval: $\varepsilon_{min}<\varepsilon_F< \varepsilon_{max}$.
Then, the Fermi surface $\varepsilon_{1,2}({\bf p})=\varepsilon_F$
of this metal in the vicinity of a point ${\bf p}_0$ defined by
\cite{c3} $\varepsilon_0({\bf p}_0)= \varepsilon_F$ consists of
two cones with the common vertex ${\bf p}_0$ and the same vertex
angle penetrated by the band-contact line, see the inset in
Fig.~\ref{fig1}. One of the cones corresponds to the band $1$, and
the second cone corresponds to the band $2$. In other words, the
Fermi surface has the self-intersecting shape. \cite{lak} At
$\varepsilon_F= \varepsilon_{min}$ this self-intersecting FS
appears, and at $\varepsilon_F= \varepsilon_{max}$ it disappears,
i.e., at these critical energies the electron topological
transitions occur, Figs.~\ref{fig1} and \ref{fig2}. These
transitions were named the appearance (disappearance) of the
self-intersecting FSs. \cite{lak}

   %%%%%%%%%%%%%%%%%%%%%%%%%%%%%%%%%%%%%%%%%%

   %%%%%%%%%%%%%%%%%%%%%%%%%%%%%%%%%%%%%%%%%%

Consider now the spectrum of the two bands near the critical point
${\bf p}_0$ at which $\varepsilon_0({\bf p}_0)=\varepsilon_c$
where $\varepsilon_c= \varepsilon_{min}$ or $\varepsilon_{max}$.
If ${\bf p}$ is measured from this ${\bf p}_0$, then we arrive at
 \begin{eqnarray}\label{3}
\varepsilon_{1,2}({\bf p})\!\!&=&\!\!\varepsilon_{c}+
B_+p_{\parallel}^2+{\bf v}_{+}^{\perp}\cdot{\bf p}_{\perp}
\\ \nonumber &\!\!\!\!\pm&\!\!\!\!\left[\left(B_-p_{\parallel}^2\!+
\!{\bf v}_-\cdot
{\bf p}_{\perp}\right)^2\!\!+ \left(B_{12}p_{\parallel}^2\!+\!{\bf
v}_{12}\cdot{\bf p}_{\perp}\right)^2 \right]^{1/2}\!\!\!\!\!\!,
 \end{eqnarray}
where $p_{\parallel}$ is the component of ${\bf p}$ parallel to
the vector ${\bf z}$, i.e., to the band-contact line at the point
${\bf p}_0$, whereas $p_{\perp}$ is the component perpendicular to
${\bf z}$. Equation (\ref{3}) takes into account that ${\bf
v}_{+}^{\parallel}=0$ at the point where $\varepsilon_0$ reaches
its extremum, and so we have included terms proportional to
$p_{\parallel}^2$ in the elements $E_{ij}$ of Hamiltonian
(\ref{1}) (the coefficient $B_+$, $B_-$, $B_{12}$ are some
constants;  $B_-$, $B_{12}$ are different from zero if the
band-contact line is not a straight line, i.e, if it is not a
symmetry axis of the crystal). Expression (\ref{3}) is the most
general form of the electron spectrum near the critical energy
$\varepsilon_c$. To analyze this spectrum, let us set the
coordinate axis $z$ along the vector ${\bf z}$, and hence the
$x$-$y$ plane coincide with the plane of the vectors ${\bf v}_-$
and ${\bf v}_{12}$. Without loss of generality, we may also assume
that the axes $p_x$ and $p_y$ are chosen so that the quadratic
form $({\bf v}_- \cdot{\bf p}_{\perp})^2+({\bf v}_{12}\cdot{\bf
p}_{\perp})^2$ under the root sign in Eq.~(\ref{3}) is diagonal
and hence takes the form: $b_{xx}p_x^2+b_{yy}p_y^2$ where
 \[
b_{xx}=(v_-)_x^2+(v_{12})_x^2,\ \ \ b_{yy}=(v_-)_y^2+(v_{12})_y^2.
 \]
Besides, at a fixed $p_z$ we shall measure $p_x$ and $p_y$ from
the band-contact line, i.e., from the point ${\bf p}_{\perp}(p_z)$
defined by the equations: ${\bf v}_{12}\cdot{\bf
p}_{\perp}+B_{12}p_{z}^2=0$, ${\bf v}_-\cdot{\bf
p}_{\perp}+B_-p_{z}^2=0$. In this coordinate system the
band-contact line is straight, $p_x=p_y=0$, and equation (\ref{3})
reduces to the formula:
 \begin{eqnarray}\label{4}
\varepsilon_{1,2}({\bf p})=\varepsilon_{c}+ Bp_{z}^2+{\bf
v}_+^{\perp}\cdot{\bf p}_{\perp}
 \pm \left[b_{xx}p_x^2+b_{yy}p_y^2 \right]^{1/2}\!\!\!\!,
 \end{eqnarray}
where
 \begin{eqnarray*}
B\!=\!{\rm det}\!\!\left (\!\begin{array}{ccc}B_+&B_-&B_{12}
\\(v_+)_x &(v_-)_x & (v_{12})_x \\ (v_+)_y & (v_-)_y & (v_{12})_y
\end{array}
\right )\!\!\!\left[{\rm det}\!\!\left(\!\!\begin{array}{cc}
(v_-)_x & (v_{12})_x
\\ (v_-)_y & (v_{12})_y
\end{array}
\!\!\right)\!\!\right]^{-1}\!\!\!\!\!\!\!\!.
 \end{eqnarray*}
It is important to emphasize that the spectrum (\ref{4}), in fact,
depends only on two essential parameters. Indeed, changing the
variables $p_x$, $p_y$, and $p_z$ as follows: $\tilde
p_x=\sqrt{b_{xx}}p_x$, $\tilde p_y=\sqrt{b_{yy}}p_y$, $\tilde
p_z=\sqrt{|B|}p_z$, and rotating the axes $\tilde p_x$, $\tilde
p_y$ through some angle, one can reduce Eq.~(\ref{4}) to the
formula:
 \begin{eqnarray}\label{5}
\varepsilon_{1,2}(\tilde{\bf p})=\varepsilon_{c}+ {\rm sgn(B)}
\tilde p_{z}^2+\tilde a \tilde p_x \pm \left[\tilde p_x^2+\tilde
p_y^2 \right]^{1/2},
 \end{eqnarray}
where ${\rm sgn}(x)=\pm 1$ for $x>0$ and  $x<0$, respectively,
and
 \begin{eqnarray}\label{6}
\tilde a^2\equiv \frac{(v_+)_x^2}{b_{xx}}+
\frac{(v_+)_y^2}{b_{yy}}.
 \end{eqnarray}
Thus, the spectrum is substantially determined by a value of
$\tilde a$ which characterizes the shape of the FS (see
Figs.~\ref{fig1}--\ref{fig3}) and by a sign of $B$ which specifies
if the self-intersecting surfaces appear or disappear at
$\varepsilon= \varepsilon_c$. The electron spectrum near any
critical point of interest can be reduced to expression (\ref{5}).

If a weak spin-orbit interaction is taken into account, inspection
of a narrow ``tube'' in the place of the band-contact line becomes
more appropriate in analyzing  the electron energy spectra. Inside
this tube the two bands closely approach each other without their
touching. Hence the intersection of the appropriate Fermi
surfaces, strictly speaking, does not occur (there is a small
space gap between them). However, as our analysis shows, this fact
has no essential effect on the results presented below, and so we,
as a rule, neglect the spin-orbit interaction here. For reference,
we note that with the spin-orbit interaction, the spectrum
(\ref{5}) transforms into,
 \begin{eqnarray}\label{5a}
\varepsilon_{1,2}(\tilde{\bf p})=\varepsilon_{c}\!+{\rm sgn(B)}
\tilde p_{z}^2+\tilde a \tilde p_x \pm \!\left[\Delta^2+\tilde
p_x^2+\tilde p_y^2 \right]^{1/2}\!\!\!\!,
 \end{eqnarray}
where $\Delta$ is the energy gap between the bands at $\tilde {\bf
p}=0$. Interestingly, at $\tilde a \neq 0$ a minimum of the upper
band and a maximum the lower band in $\tilde{\bf p}_{\perp}$ occur
at different values of $\tilde{\bf p}_{\perp}$, and the spectrum
(\ref{5a}) is characterized by the indirect energy gap
$2\Delta_{min}=2\Delta(1-\tilde a^2)^{1/2}$. At $\tilde a^2>1$ the
indirect gap is zero.

\section{Electron topological transitions of $3\frac{1}{2}$ kind}

Near $\varepsilon_F=\varepsilon_{c}$ the density of the electron
states $\nu(\varepsilon_F)$ calculated per unit volume of the
crystal is the sum of its regular part $\nu_{reg}(\varepsilon_F)$
and of its singular part $\delta \nu(\varepsilon_F)$. The latter
is due to the electron topological transitions mentioned above and
differs from zero when $\Delta\varepsilon_F \equiv (\varepsilon_F-
\varepsilon_{c}){\rm sgn}(B)>0$. This singular $\delta
\nu(\varepsilon_F)$ exists only if $\tilde a^2<1$,
 \begin{eqnarray}\label{7}
\delta \nu(\varepsilon_F)=\frac{4 (\Delta\varepsilon_F)^{3/2}}
{3\pi^2\hbar^3|B|^{1/2}(1-\tilde a^2)^{3/2}(b_{xx}b_{yy})^{1/2}},
 \end{eqnarray}
where we have taken into account the double degeneracy of the
electron states in spin. At $\tilde a=0$ formula (\ref{7})
reproduces the density of the electron states near a critical
energy in wurzite type crystals. \cite{rashba,lak}  If $\tilde
a^2>1$, the singular term is absent. Note that the singularity in
the density of the electron states, $\delta \nu\propto
(\varepsilon_F- \varepsilon_{c})^{3/2}$, is weaker than the
singularity $\delta \nu\propto (\varepsilon_F-
\varepsilon_{c})^{1/2}$ at the well-known topological transitions
of $2\frac{1}{2}$ kind \cite{lak,lif,var,var1} when a
Fermi-surface void appears (disappears) or a neck of the Fermi
surface forms (disrupts). It also follows from formula (\ref{7})
that the singular contribution to the $\Omega$-potential has now
the form, $\delta \Omega(\varepsilon_F)\propto (\varepsilon_F
-\varepsilon_c)^{7/2}$, and  the topological transitions
considered here are the $3\frac{1}{2}$-order transitions according
to the classification of Lifshits. \cite{lif,lak}

\begin{figure}[t] % %%%%%%%%%%%%%%%%%%%%%%%%%%%%%%%%%%%%%
 \centering  \vspace{+9 pt}
\includegraphics[scale=.80]{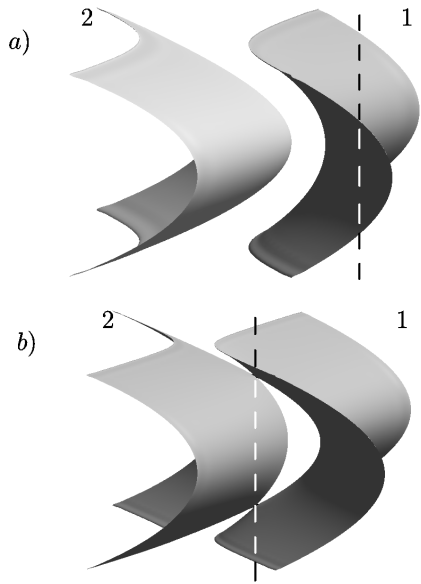}
\caption{\label{fig3} The Fermi surfaces near
$\varepsilon_c=\varepsilon_{min}$ at $\tilde a^2>1$: (a)
$\varepsilon_F < \varepsilon_{min}$, (b) $\varepsilon_F >
\varepsilon_{min}$. The band-contact line is shown by the dash
line.
 } \end{figure}

It is known \cite{top} that the topological properties of any
closed surface are completely determined by the number of its
components (its disconnected parts) and by the number of handles
for each of the components. How can the transitions consisting in
appearing (disappearing) the self-intersecting Fermi surfaces be
understood from this point of view? (Recall that, in reality, there
is a small gap between the touching FSs if one takes into account
a weak spin-orbit interaction, and the intersection of the
surfaces, strictly speaking, does not occur.) It is clear from
Fig.~\ref{fig1} that at $\tilde a < 1$ the appearance of the
self-intersecting FS is equivalent to the disappearance of the
handle that exists in the FS of the first band at $\varepsilon_F<
\varepsilon_{min}$ and to the simultaneous appearance of the
ovaloid of the second band. The disappearance of the handle can be
also interpreted as the disruption of the Fermi-surface neck, and
the choice of the interpretation depends on the shape of the FS
for the first band far away from the critical point. In
Fig.~\ref{fig2} we show the disappearance of the self-intersecting
FS in the vicinity of $\varepsilon_{max}$ at $\tilde a < 1$. This
transition is equivalent to the disappearance of the void of the
first band and to the formation of the Fermi-surface neck of the
second band (or to the appearance of the handle in another
interpretation). For these two combined transitions at
$\varepsilon_{min}$ and $\varepsilon_{max}$, the changes of the
Fermi-surface components or their handles are evident. On the
other hand, at $\tilde a > 1$ the appearance of the
self-intersecting FS does not change the number of its components
and handles, Fig.~\ref{fig3}. In other words, there is no
topological transition at $\tilde a > 1$, and it is for this
reason that the density of the electron states does not exhibit
any singularity in this case. In this context, it would be more
properly to name the topological transitions discussed here the
combined electron topological transitions rather than the
appearance (disappearance) of the self-intersecting Fermi
surfaces.

   %%%%%%%%%%%%%%%%%%%%%%%%%%%%%%%%%%%%%%%%%%

At the combined topological transitions of the $3\frac{1}{2}$ kind
the singularities of the physical quantities that are proportional
to the density of the electron states $\nu$ or its derivative
$d\nu/d\varepsilon_F$ are weaker than the singularities at the
topological transitions of the $2\frac{1}{2}$ kind. This makes a
detection of the $3\frac{1}{2}$-order transitions difficult with
measurements of these quantities. However, these transitions can
be detected with the magnetic susceptibility since its orbital
part generally is not determined by the density of the electron
states, and at the $3\frac{1}{2}$-order transitions the component
$\chi_{zz}$ of the magnetic-susceptibility tensor turns out to
exhibit a giant diamagnetic anomaly. \cite{m-sv} At low
temperatures $T$, $T\ll \Delta\varepsilon_F$, and weak magnetic
fields $H$, $H\ll H_T\equiv
T^2c/[e\hbar(b_{xx}b_{yy})^{1/2}(1-\tilde a^2)^{3/2}]$, one has
\cite{m-sv}
 \begin{eqnarray}\label{8}
\chi_{zz}^0(\varepsilon_F)=-\frac{e^2}{6\pi^2 c^2\hbar}\left (
\frac{b_{xx}b_{yy}}{|B|}\right)^{1/2}\frac{(1-\tilde
a^2)^{3/2}}{(\Delta \varepsilon_F)^{1/2}}.
 \end{eqnarray}
Thus, the anomaly is the large peak in the
$\varepsilon_F$-dependence  of $|\chi_{zz}^0|$. Interestingly, the
singularity in the susceptibility, $\chi_{zz}^0(\varepsilon_F)
\propto (\Delta \varepsilon_F)^{-1/2}$, is the same as the
singularity in the thermoelectric power at the
$2\frac{1}{2}$-order transitions. \cite{var} The divergence of
expression (\ref{8}) for $\Delta \varepsilon_F \to 0$ should be
cut off at $\Delta \varepsilon_F \sim T$, and this cut-off
determines the magnitude of the diamagnetic peak at the weak
magnetic fields.

In the case $H> H_T$ the magnetic susceptibility near the points
of $3\frac{1}{2}$-order transitions were theoretically studied in
detail in Ref.~\onlinecite{m-sh} using the exact expression for
the electron Landau levels in the vicinity of these points at the
magnetic field ${\bf H}\parallel {\bf z}$,
 \begin{eqnarray}\label{8a}
\varepsilon_{1,2}(l,\tilde p_z)=\varepsilon_{c}\!+{\rm sgn(B)}
\tilde p_{z}^2 \pm \!\left[\frac{eH\alpha}{c\hbar}l
+\Delta_{min}^2\right]^{1/2}\!\!\!\!,
 \end{eqnarray}
where $\alpha=2(b_{xx}b_{yy})^{1/2}(1-\tilde a^2)^{3/2}$; $\tilde
a^2<1$;  $l$ is a non-negative integer ($l=0$, $1$, \dots), and
$\Delta_{min}=\Delta(1-\tilde a^2)^{1/2}$. At $\tilde a^2>1$ the
shape of the FS near the touching points (see Fig.~\ref{fig3})
does not lead to the discrete spectrum in the magnetic field, and
formula (\ref{8a}) fails. With spectrum (\ref{8a}), we obtain at
high magnetic fields $H\gg \Bar{H} \equiv {\rm max}(H_T,H_T(\Delta
\varepsilon_F)^2 /T^2,H_T(\Delta_{min})^2/T^2$), \cite{m-sh}
\begin{eqnarray}\label{9}
 \chi_{zz}(H)=-\frac{3\cdot
f_0}{2^{5/4}\pi^2}\frac{e^{7/4} (b_{xx}b_{yy})^{3/8}(1-\tilde
a^2)^{9/8} }{c^{7/4}\hbar^{5/4}|B|^{1/2}H^{1/4}},
 \end{eqnarray}
where $f_0\approx 0.156$. Expression (\ref{9}) can be also
rewritten in the form $\chi_{zz}(H)= (9f_0/\sqrt
2)\chi_{zz}^0(\varepsilon_F) (\Bar{H}/H)^{1/4}$ which reveals the
$H$-dependence of the diamagnetic peak magnitude and shows that
the the giant anomaly is suppressed  in strong magnetic fields.

Finally, it is worth noting that the band-contact lines in
crystals can, in principle, intersect each other, and at the
points of their intersection more exotic electron topological
transitions than the $3\frac{1}{2}$-order transitions considered
here are possible. At some of these exotic transitions the
magnetic susceptibility can exhibit unusual anomalies. \cite{mik}

\section{Conclusions}

We have theoretically studied the electron topological
transitions of the $3\frac{1}{2}$ kind associated with
band-contact lines in 3D metals. In the 3D metals with the
inversion symmetry and a weak spin-orbit interaction these
transitions are widespread no less than the well-known
$2\frac{1}{2}$-order transitions of Lifshits.


\begin{thebibliography}{}
\bibitem{c3} There are at least two such points in the Brillouin
zone.
\end{thebibliography}
\end{document}